\documentclass[aps,prl,superscriptaddress,showpacs,preprint]{revtex4}

\usepackage{graphicx,amssymb,amsmath}
\usepackage[]{color}
\usepackage{ulem}

\begin{document}


\title{Controlling the 2DEG states evolution at a metal/Bi$_2$Se$_3$ interface}


\author{Han-Jin Noh}
\author{Jinwon Jeong}
\author{En-Jin Cho}
\affiliation{Department of Physics, Chonnam National University, Gwangju 500-757, Korea}
\author{Joonbum Park}
\author{Jun Sung Kim}
\affiliation{Department of Physics, Pohang University of Science and Technology, Pohang 790-784, Korea}
\author{Ilyou Kim}
\author{Byeong-Gyu Park}
\affiliation{Pohang Accelerator Laboratory, Pohang University of Science and Technology, Pohang 790-784, Korea}
\author{Hyeong-Do Kim}
\affiliation{Center for Correlated Electron Systems, Institute for Basic Science (IBS), Seoul 151-747, Korea}
\affiliation{Department of Physics and Astronomy, Seoul National University, Seoul 151-747, Korea}


\date{\today}

\begin{abstract}
We have demonstrated that the evolution of the two-dimensional electron gas (2DEG) system at an interface of metal and the model topological insulator (TI) Bi$_2$Se$_3$ can be controlled by choosing an appropriate kind of metal elements and by applying a low temperature evaporation procedure.
In particular, we have found that only topological surface states (TSSs) can exist at a Mn/Bi$_2$Se$_3$ interface, which would be useful for implementing an electric contact with surface current channels only.
The existence of the TSSs alone at the interface was confirmed by angle-resolved photoemission spectroscopy (ARPES).
Based on the ARPES and core-level x-ray photoemission spectroscopy measurements, we propose a cation intercalation model to explain our findings.

\vspace{20mm}
*E-mail: ffnhj@chonnam.ac.kr

\end{abstract}


\maketitle

Since the discovery of topological insulators (TIs) with topologically protected metallic surface states\cite{Fu1, Noh1, Hsieh1, Xia, HZhang, Hsieh2}, much attention has been paid to characterization of the topological surface states (TSSs), unveiling the abundant exotic properties of the TSSs such as the absence of back-scattering\cite{Roushan, TZhang}, the spin-momentum locking\cite{Hsieh2, Pan}, the robustness against various kinds of surface perturbations\cite{Park1, Noh2, Valla1, Wray}.
In particular, the studies on the effect of surface adsorbates to the TSSs have found very intriguing phenomena that new Rashiba-type spin-split surface states emerge on the (111) surface of Bi$_2$Se$_3$ TI and that they have large amounts of common features irrespective of the electric/magnetic properties of the adsorbates\cite{Valla1, King, Benia}.
The successive experimental and theoretical studies have suggested that the newly emerging surface states are actually interface states in a quantum-confined two-dimensional electron gas (2DEG) induced by a surface band-bending effect\cite{Bianchi, Bahramy, Mann}.

The coexistence of TSSs and 2DEG states is a very exotic example in a viewpoint of surface physics and have potential applications in the field of nanoscale spintronic devices, so the understanding of their origin and properties is essential in this TI physics.
However, the newly developed 2DEG has a complex spin-orbit-split electronic structure, which could be an obstacle depending on the way how to implement TI-based devices.
For example, when we want to use an electric current through only the TSSs in Bi$_2$Se$_3$ or when we want to make an n-type TI-metal junction whose current channel is only TSSs, those 2DEG states are definitely to be removed.
In spite of these practical needs, all reports on the effects of adsorbates or surface impurities on Bi$_2$Se$_3$ surfaces have only shown that the 2DEG states are easily formed by various kinds of adsorbates in an ultra-high vacuum, but have not suggested a possible way to remove the developed 2DEG states without killing the TSSs.

In this study, we focus on this issue and performed systematic angle-resolved photoemission spectroscopy (ARPES) measurements in order to explore whether there is a way to remove the 2DEG states with leaving the TSSs alone at a metal/Bi$_{2}$Se$_{3}$ interface.
Keeping in mind that the practicality of the surface states described above appears at an interface between metal and TI, we deposited {\it in situ} several kinds of metal elements such as Cu, In, and Mn on a (111) surface of Bi$_{2}$Se$_{3}$ in a well-controlled manner, which actually corresponds to an early stage of a metal thin film synthesis, and observed an evolution/devolution of the TSSs and 2DEG states as a function of deposition thickness by ARPES and core-level x-ray photoemission spectroscopy (XPS).
Our ARPES measurements reveal that in the case of Cu and In, the 2DEG states are developed with a Rashiba-type spin-split electronic structure and are saturated up to $\sim$1 monolayer (ML), but that in the case of Mn, the 2DEG states are developed up to $\sim$0.4 ML, then disappear above $\sim$1.0 ML.
The disappearance of the 2DEG states and the existence of the TSSs alone were observed on a Bi$_2$Se$_3$ (111) surface covered with up to $\sim$10 ML of Mn atoms by ARPES, indicating that only the TSSs exist at the Mn/Bi$_2$Se$_3$ interface prepared by our procedure.
In order to explain why Mn deposition makes a difference from Cu or In deposition, we kept track of the XPS spectral changes for Bi 4$f$, Se 4$d$, and Cu/In/Mn core levels, and proposed a cation intercalation model where the intercalated cations act as a potential gradient reducer.

\subsection{Results}

\begin{figure*}
\includegraphics[width=12.0 cm]{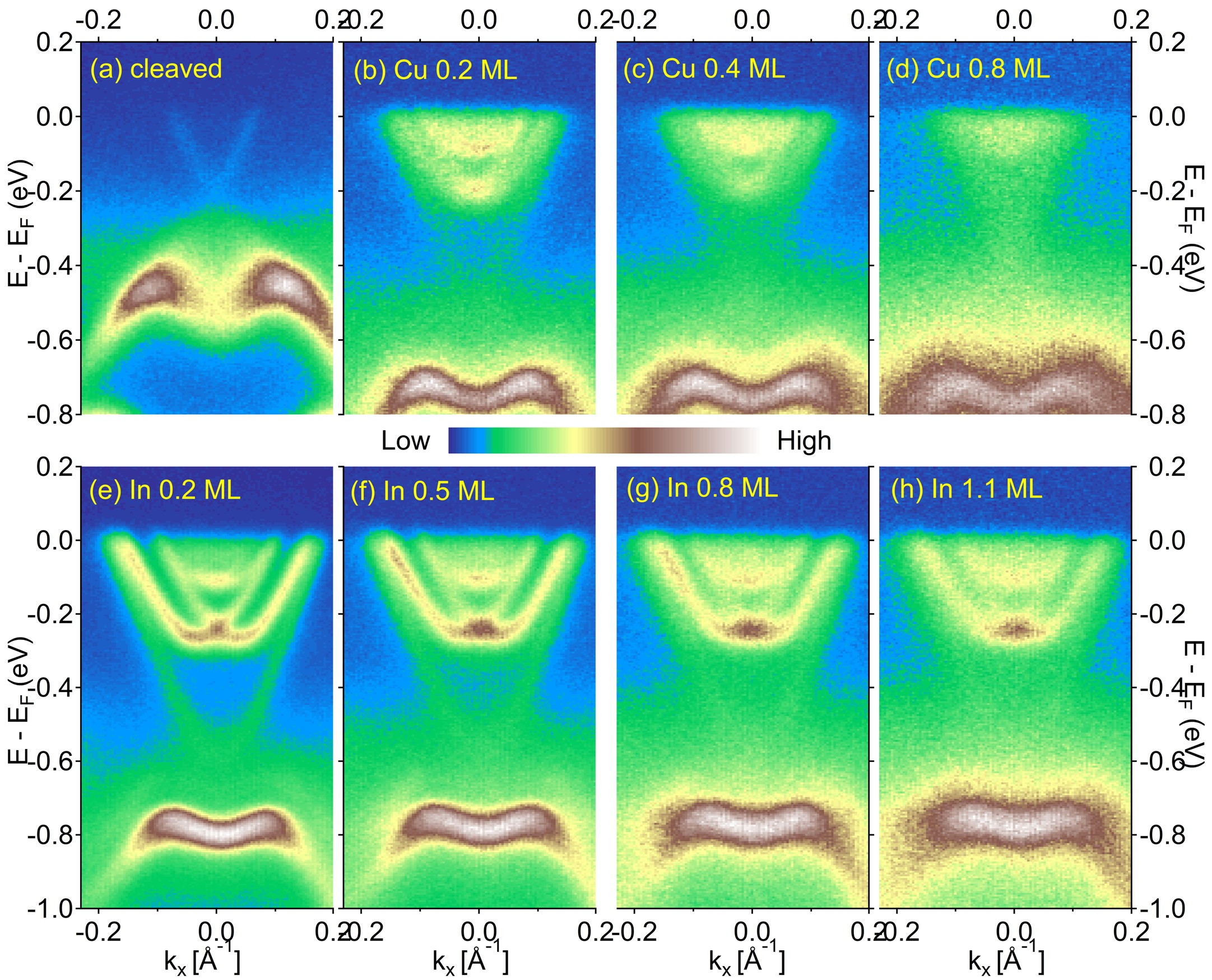}
\caption{\label{CuIn} Evolution of the TSSs and the 2DEG states on the (111) surface of Bi$_2$Se$_3$ TI as a function of Cu/In deposition thickness.
(a) ARPES-measured TSSs obtained from a clean (111) surface of Bi$_2$Se$_3$ along the $\overline{\Gamma}-\overline{K}$ direction within 15 minutes after {\it in situ} cleaving at 40 K. (b)-(d) ARPES images after Cu 0.2, 0.4, and 0.8 ML deposition, respectively. (e)-(h) ARPES images after In 0.2, 0.5, 0.8, and 1.1 ML deposition, respectively.
}
\end{figure*}

Typical features of the TSSs in a fresh (111) surface of Bi$_2$Se$_3$ are presented in Fig.~\ref{CuIn}(a).
The ARPES image was obtained along the $\overline{\Gamma}-\overline{K}$ direction within 15 minutes after cleaving the sample at 40 K during which we did not observe any aging effect due to the adsorption of residual gas molecules in the vacuum chamber.
This is quite different from the p-doped Bi$_2$Te$_3$ case in our previous report\cite{Noh1}.
The V-shaped linear dispersive surface bands form two Dirac cones sharing the Dirac point (DP) located at around 200 meV below the Fermi level.
Below the DP, a clear M-shaped bulk valence band is seen.
On this cleaved surface, we deposited copper or indium atoms {\it in situ} by evaporating pure (99.999\%) copper or indium metal with a well-controlled manner described below.
Figures~\ref{CuIn}(b)-\ref{CuIn}(d) show an evolution of the TSSs and the 2DEG states as a function of Cu deposition thickness.
At a small amount of Cu deposition ($\sim$0.2 ML), newly developed surface states with Rashiba-type spin-splittings are clearly seen together with the TSSs.
The DP and the M-shaped valence bands shift to the higher binding energy side by $\sim$0.3 eV.
In Figs.~\ref{CuIn}(e)-\ref{CuIn}(h), a similar surface state evolution is displayed with the increase of In deposition.
A prominent difference from the Cu deposition case appears in the size of the Rashba-type spin splittings, but both cases share common features in many aspects.
The similar behavior of the surface state evolution has been reported for various kinds of surface deposition or surface adsorption\cite{Benia, Valla1}.
However, the development of the surface states does not originate from a topological property of TIs but from an interface property of metal/semiconductor.
In the case of Bi$_{0.9}$Sb$_{0.1}$ TI, our previous ARPES study shows just a small shift of the Fermi level in the electronic structure with response to the surface adsorption\cite{Noh2}.

The origin of these surface states emerging in the (111) surface of Bi$_2$X$_3$ (X=chalcogen) compounds has proved controversial initially, but one of the most persuasive models argues that those are a kind of 2DEG states developed in a 2-dimensional quantum well which is induced by a strong band bending at a metal/semiconductor interface\cite{Noh1, Wray, King, Bianchi, Bahramy}.
According to this scenario, the dominant factors for the 2DEG states are the shape of the potential profile as a function of depth from the interface and the induced charge density at the interface, but the kinds and the amount of adsorbates are not important if only they stick on the surface to form a well-defined interface.
Thus, in order to remove the 2DEG states at a metal/Bi$_2$Se$_3$ interface, or at least in order to make a different surface electronic structure from the prototypical 2DEG states, metal elements that can intercalate or be interstitial defects are expected to be more effective.
In this viewpoint, one different behavior between Cu- and In-depositions can be qualitatively understood.
At relatively thick deposition ($\gtrsim$0.8 ML), the 2DEG states look more shrunk in the Cu-deposited sample than in the In-deposited sample as shown in Figs.~\ref{CuIn}(d) and \ref{CuIn}(g).
This is possibly due to the intercalatability difference between Cu and In atoms for Bi$_2$Se$_3$.
Actually, it has been reported that Cu atoms can intercalate into the van der Waals (vdW) gaps in Bi$_2$Se$_3$, while there is no such report for In atoms to our knowlege\cite{Wang}.

\begin{figure*}
\includegraphics[width=12.0 cm]{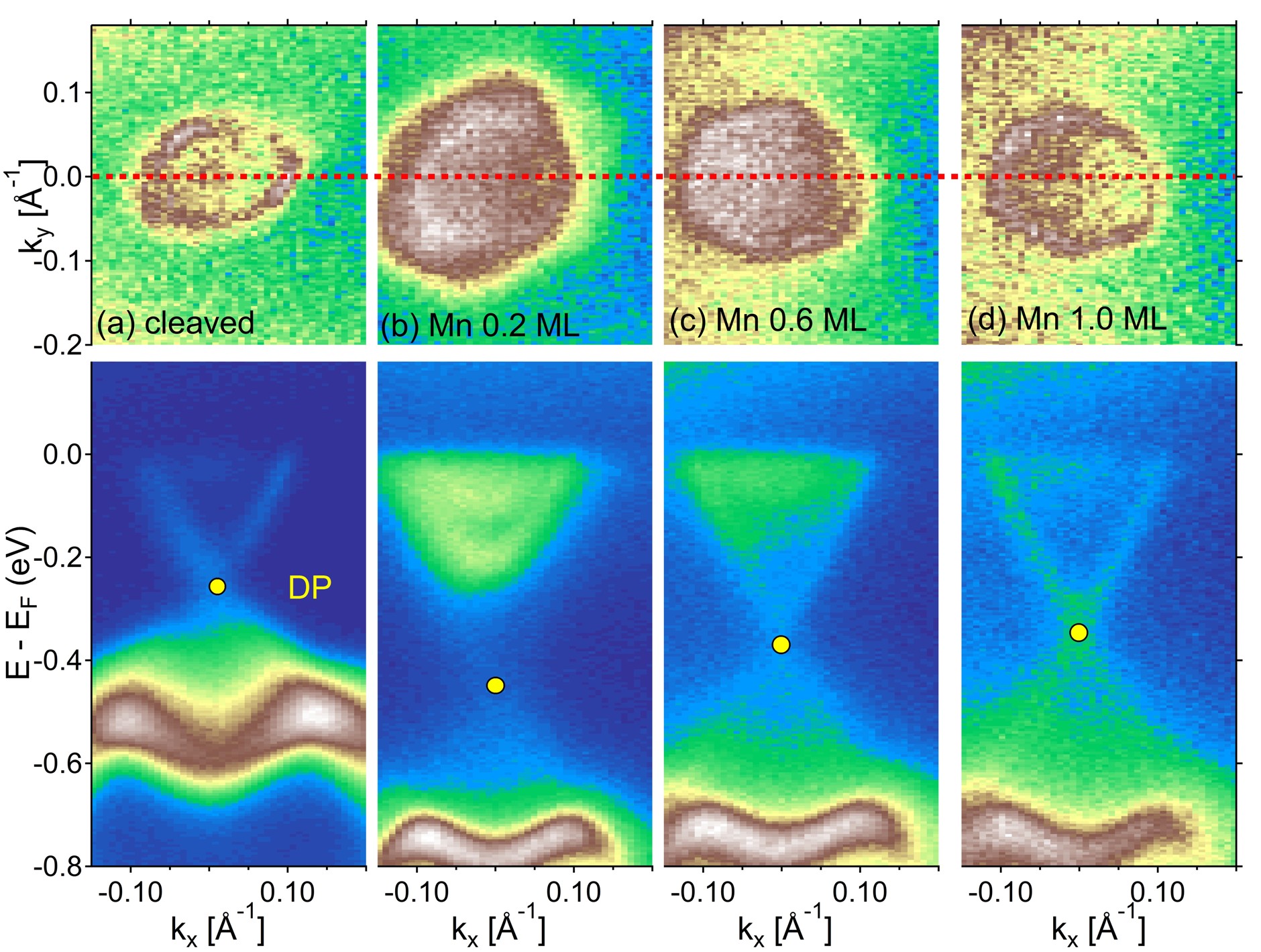}
\caption{\label{Mn}
Evolution of the TSSs and the 2DEG states on the (111) surface of Bi$_2$Se$_3$ TI as a function of  Mn deposition thickness.
(a) ARPES-measured FSs (upper) and the dispersion relation of TSSs (lower) obtained from a clean (111) surface of Bi$_2$Se$_3$ along the $\overline{\Gamma}-\overline{M}$ direction within one hour after {\it in situ} cleaving at 40 K. (b)-(d) The corresponding ARPES images after Mn 0.2, 0.6, and 1.0 ML deposition, respectively. Up to $\sim$0.2 ML of Mn, the 2DEG states are developed, but they becomes weaker and the Dirac point moves $\sim$0.1 eV upward at Mn $\sim$0.6 ML thickness. The 2DEG states almost disappear at $\sim$1.0 ML thickness, and only TSSs are seen in the ARPES images.
}
\end{figure*}

Keeping in mind that the intercalatability of the deposited atoms may be the crucial factor for our purpose, we chose Mn as a deposition material and kept track of the 2DEG states evolution as a function of deposition thickness since it is known that a small amount of Mn can be doped in Bi$_2$Se$_3$\cite{Cho}.
Rubidium was also reported to intercalate into the gap, but its effect on the 2DEG states lies in another direction to our purpose\cite{Bianchi2}.
Figures~\ref{Mn}(a)-\ref{Mn}(d) show the ARPES-measured Fermi surface (upper panel) and the corresponding energy dispersion relation (lower panel) along the red dotted line (approximately $\overline{\Gamma}-\overline{M}$ direction) for a clean surface, 0.2 ML, 0.6 ML, and 1.0 ML Mn-deposited surface, respectively.
Interestingly enough, Mn-deposited surfaces show a very different evolution of the 2DEG states.
At a small amount of Mn deposition ($\sim$0.2 ML), quite a complex 2DEG structure is developed as shown in Fig.~\ref{Mn}(b).
It looks similar to that of the Cu-deposited surface.
However, when the deposited layer is thicker than $\sim$0.2 ML, the Fermi surface (FS) size of the 2DEG states becomes smaller, and the DP shifts up by $\sim$0.1 eV as is shown in Fig.~\ref{Mn}(c).
At around Mn 1.0 ML, the 2DEG states almost disappear and only the TSSs remain as can be seen in Fig.~\ref{Mn}(d).
Based on the facts that the energy position of the DP in Mn 1.0 ML deposition is $\sim$0.1 eV deeper than that of the clean surface and that the size of the hexagonal FS of the TSSs is a little larger than that of the clean surface, the surface does not recover to its original condition, but forms another interface that gives a similar environment to the original vacuum/(111) interface for the TSSs and 2DEG.

\begin{figure*}
\includegraphics[width=12.0 cm]{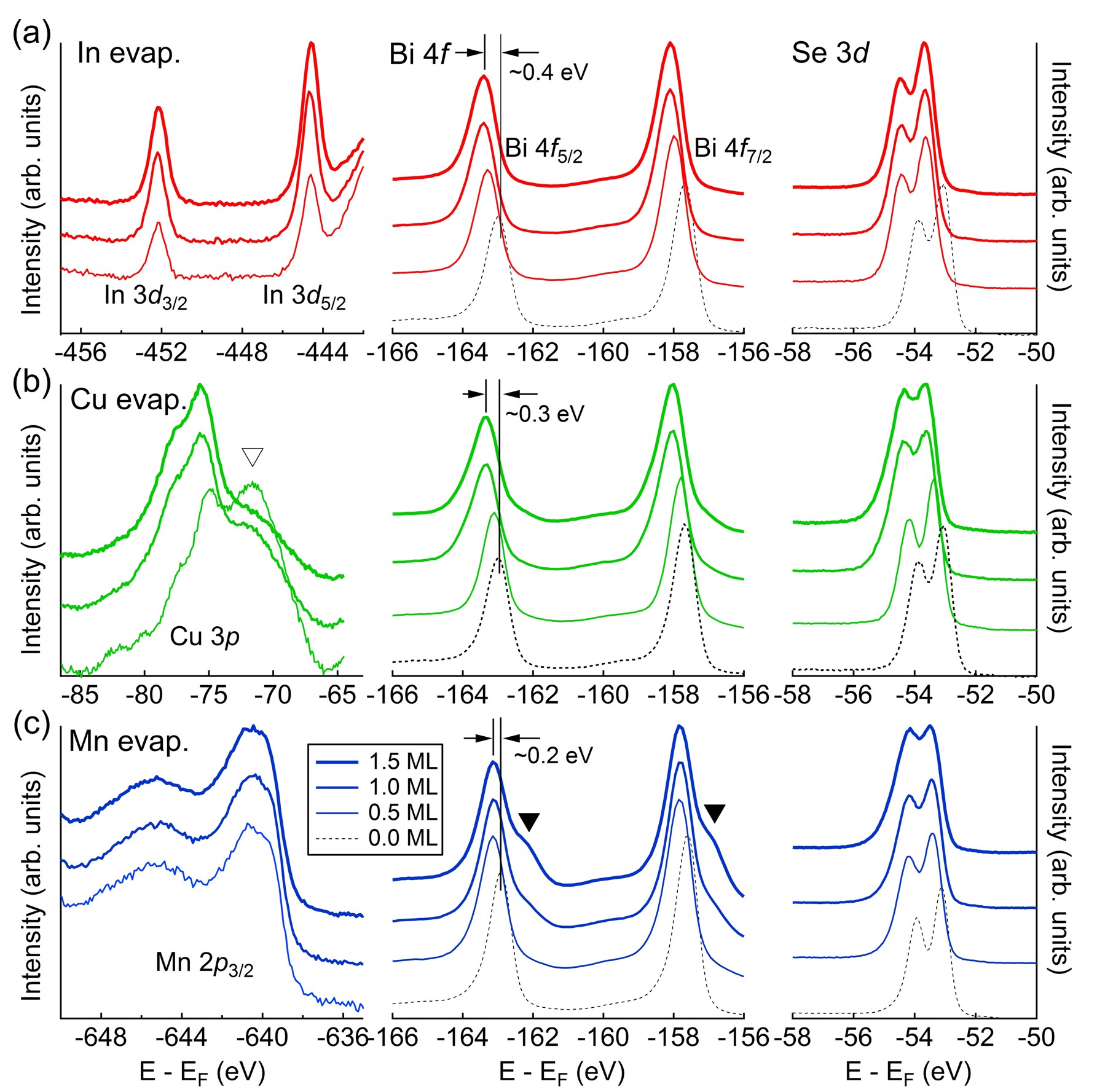}
\caption{\label{XPS}
Core level XPS spectra of the deposited element (left column), Bi 4$f$ (center column), and Se 3$d$ (right column) on the (111) surface of Bi$_2$Se$_3$ with the increase of deposition thickness. The line width of the spectra indicates the deposition thickness.
(a) The In 3$d$ peak position does not change with the In thickness, while the Bi 4$f$ and Se 3$d$ peak positions get lower by $\sim$0.4 eV.
(b) The Cu 3$p$ peaks show two doublet structures, each of which corresponds to surface Cu and intercalated Cu ($\bigtriangledown$), respectively. The intercalated Cu peak gets relatively weaker as the Cu deposition thickness increases.
(c) The Mn 2$p_{3/2}$ peak hardly depends on the Mn deposition thickness, but Bi 4$f$ and Se 3$d$ peaks show an energy shift by $\sim$0.2 eV and a satellite peak evolution($\blacktriangledown$). The satellite peak in Bi 4$f$ indicates the existence of the interstitial Mn cations.
}
\end{figure*}

In order to figure out why the Mn deposition makes such an intriguing behavior in the 2DEG states evolution, we carried out core-level XPS measurements as is displayed in Fig.~\ref{XPS}.
Since XPS is sensitive to the chemical valency and chemical environment of ions in solids, important information or at least a clue on the reason of the disappearance of the 2DEG states can be obtained by analyzing the XPS spectra.
In the left column, the center one, and the right one of Fig.~\ref{XPS}, the XPS spectra of each deposited element core level, Bi 4$f$, and Se 3$d$ level are presented with the increase of the deposition thickness, respectively.
In the case of In deposition, the Bi 4$f_{5/2,7/2}$ and Se 3$d_{3/2,5/2}$ peaks shift toward lower kinetic energy side by $\sim$0.4 eV with the increase of the deposition thickness as shown in Fig.~\ref{XPS}(a).
The amount of the energy shift corresponds to the band bending shown in Figs.~\ref{CuIn}(a) and \ref{CuIn}(h), so it can be interpreted as a surface potential shift due to the accumulated surface charge induced by the In deposites.
The similar core-level shifts are observed in the Cu- and Mn-deposited surface, respectively, as is shown in Figs.~\ref{XPS}(b) and \ref{XPS}(c), and the amount of energy shift is also very consistent with that of the band shift in Figs.~\ref{CuIn} and \ref{Mn}.

Several interesting features in the core-level spectra of each deposited element, Bi 4$f$, and Se 3$d$ are found in a different dependence on deposition.
While In deposition does not make any change in the In 3$d$, Bi 4$f$, and Se 3$d$ spectral line shape, Cu(Mn) deposition induces an evolution in Cu 3$p$ (Mn 2$p$), Bi 4$f$, and Se 3$d$ level, respectively.
In the Cu 3$p$ region, two doublets (75 eV and 71 eV) are observed, each of which can be assigned to surface peak and intercalation peak($\bigtriangledown$) as shown in Fig.~\ref{XPS}(b).
The relative peak weight dependence on deposition thickness strongly supports this assignment, and the similar result has been reported in the Rb deposition/annealing study\cite{Bianchi2}.
Meanwhile, Mn 2$p_{3/2}$ do not make a prominent difference in their shape with the amount of Mn deposition, so it is not clear whether a part of the adsorbated Cu atoms intercalates or not.
However, Mn deposition effects on the other core level spectra are not similar to the indium case.
For both Cu and Mn deposition cases, a new spectral weight appears between the spin-orbit-split peaks of the Se 3$d$ together with the increase of the peak width.
This indicates an occurrence of another kind of Se ions chemically different from those in pristine Bi$_2$Se$_3$.
The similar evolutionary behavior is observed in the Bi 4$f$ region ($\blacktriangledown$) of Fig.~\ref{XPS}(c).
The prominent evolution of the Bi 4$f$ peaks in the Mn-deposited surface is quite contrastive to the Cu case where the Bi 4$f$ peaks hardly show the dependence on the Cu deposition thickness.
This element-specific response to the deposites indicates that intercalatability and intercalated sites of the deposites into Bi$_2$Se$_3$ bulk are different among In, Cu and Mn.
The invariance of the XPS spectra in the In-deposited surface suggests that In atoms do not intercalate into the Bi$_2$Se$_3$ bulk.
Meanwhile, a part of the Cu or Mn deposites definitely intercalate into the bulk but their intercalated sites are different.
Based on the Bi 4$f$ and Se 3$d$ XPS spectra, the intercalated Cu atoms affect on the Se anions only.
Meanwhile, the intercalated Mn atoms affect on the Bi ions as well as the Se ions.
This contrastive response suggests that the intercalated Cu atoms reside mainly in the vdW gaps between the quintuple layers (QLs) while the Mn atoms are in the interstitial sites as well as in the vdW gaps.
This also explains why the binding energy of the intercalated Cu atom 3$p$ levels is smaller than that of the deposited Cu atoms on the surface.
Since the QLs are chemically very stable and electrically close to neutral, the intercalated atoms into the vdW gaps are also very close to charge-neutral, so the binding energy of the atoms is smaller than that of Cu metal which is close to monovalent ion.
The zero valence of the intercalated Cu ions into the vdW gaps of Bi$_2$Se$_3$ has been reported in an electron energy loss spectroscopy study\cite{Koski}.
Meanwhile, the interstitial Mn atoms are likely to act as cations as Bi atoms do.
If this is the case, the intercalated Mn cations cannot help making moderate the surface band bending effect induced by the deposited Mn adsorbates.

\subsection{Discussion}

\begin{figure*}
\includegraphics[width=12.0 cm]{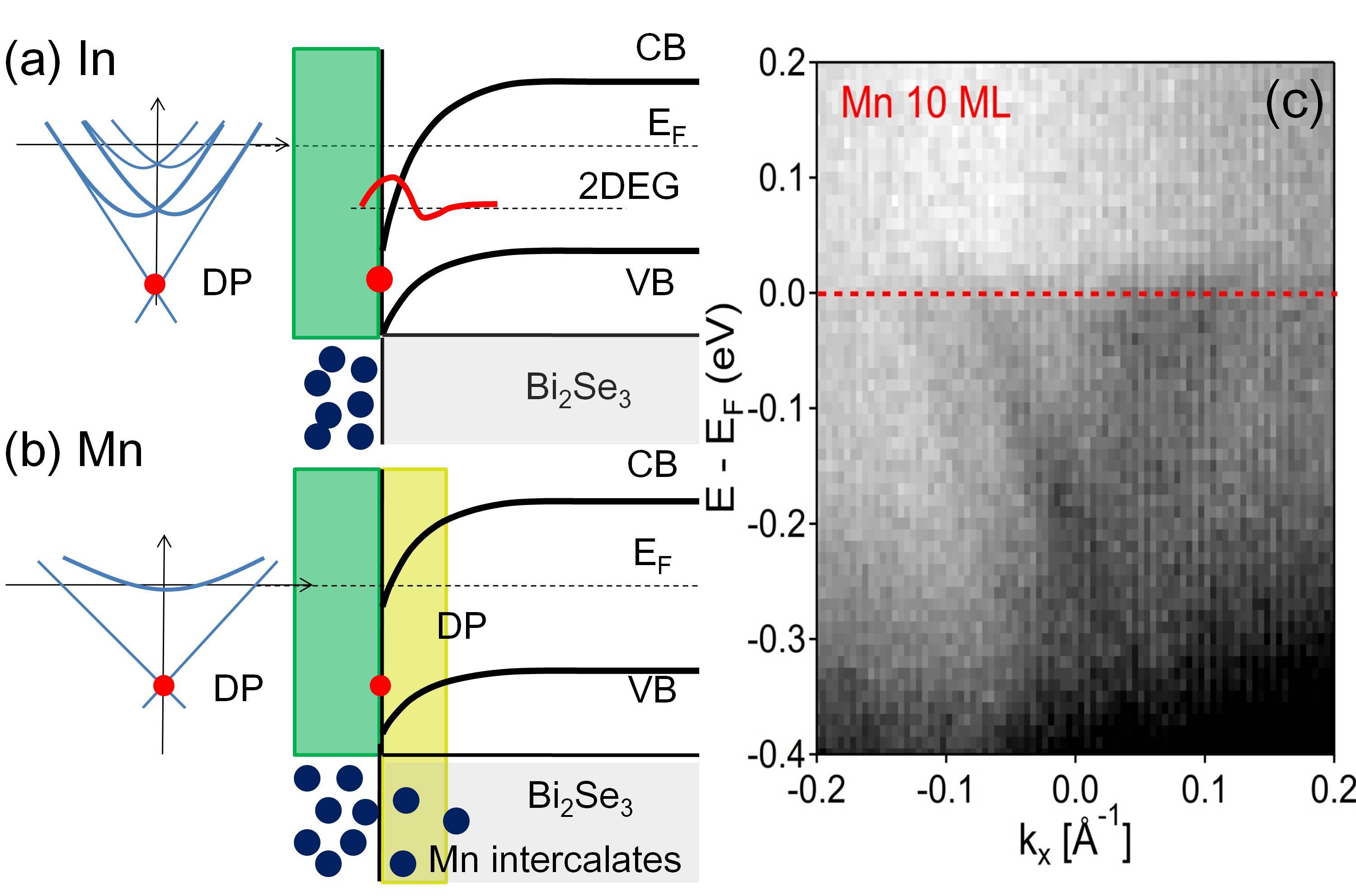}
\caption{\label{MTI}
Schematics for the surface electronic structure of a metal/Bi$_2$Se$_3$ interface.
(a) In the case of In deposition, the atoms are adsorbed on the surface of Bi$_2$Se$_3$, which induces a strong downward band bending at the metal/TI interface.
(b) In the case of Mn deposition, the Mn atoms intercalate presumably into the interstitial sites as well as the vdW gaps, which reduces the potential gradient beneath the interface.
(c) The ARPES image after Mn 10 ML deposition. Even when the Mn-deposited layers are $\sim$10 ML thick, there are only TSSs at the interface.
}
\end{figure*}

The XPS measurements described above give important information to explain the 2DEG states appearance/disappearance in our ARPES data.
Probably, the most simple way to the 2DEG removal is to weaken the adsorbates-induced band bending at the interface.
We have evidenced in the XPS study that the Mn adsorbates intercalate into the interstitial sites as well as the vdW gap between the QLs of Bi$_2$Se$_3$ and make the surface potential gradient more gentle, acting as a buffer.
Although most of the intercalated Mn atoms stay charge neutral, a part of them  gives a few electrons per atom and act as Mn cations, so the surface potential gradient gets smaller as the average intercalation depth gets larger.
Figures~\ref{MTI}(a) and \ref{MTI}(b) show a schematic configuration for our cation intercalation model.
In the case of In deposition, the surface charges induced by the deposited indium ions attract electrons toward the In/Bi$_2$Se$_3$ interface, making a 2DEG system.
When there are no intercalating indium ions, a deep well-like surface potential forms at the interface, developing 2DEG states with a complex Rashba-type spin-splitting structure.
However, as in the case of Mn deposition, if parts of the deposited Mn atoms intercalate into the interstitial sites of Bi$_2$Se$_3$, and are ionized into Mn cations, the deep well-like surface potential changes to a wide valley with a gentle slope.
In this model, the different response to Cu deposition is also naturally explained.
As is described above, Cu atoms are also known to intercalate into the vdW gap, but they do not intercalate into the interstitial sites, so almost all of the intercalated Cu atoms stay charge neutral.
The lower binding energy of the intercalated Cu 3$p$ peak in Fig.~\ref{XPS}(b) supports this interpretation.
Actually, if we compare the Cu and In deposition results, the 2DEG states in the Cu deposition shrink a little more than those in the In deposition (See Fig.~\ref{CuIn}(d) and (g)).
So, the Cu intercalation effects definitely exist, but are not so prominent like the Mn case.
This is due to a smaller fraction of the ionized Cu atoms to the intercalated ones than that of Mn atoms.

Finally, we have checked whether the TSSs still exist even in an environment that can be regarded as a real metal/TI interface.
In Fig.~\ref{MTI}(c), the ARPES image shows dim V-shape TSSs near the Fermi level.
This image was obtained from a 10 ML Mn-deposited Bi$_2$Se$_3$ (111) surface.
Since the scattering rate of the photoelectrons is very high due to the thick Mn layers, the ARPES-measured TSSs look very blurred, but definitely we can see the existence of the TSSs alone at the interface.

\subsection{Experimental methods}

The Bi$_2$Se$_3$ single crystals were grown by the melting method. The stoichiometric mixture of Bi(99.999$\%$) and Se(99.9999$\%$) was loaded in a evacuated quartz ampoule. The ampoule was heated up to 850$^{\circ}$C for 48 hours, followed by slow cooling to 500$^{\circ}$C at a rate of 2$^{\circ}$C/h. The furnace was kept at this temperature for 5 additional days for post-annealing before furnace cooling.
The ARPES and XPS experiments were performed at the 4A1 beamline of the Pohang Light Source with a Scienta SES-2002 electron spectrometer and $\hbar\omega$=25 and 800~eV photons\cite{HDKim}.
The total energy (momentum) resolution is $\sim$20~meV ($\sim$0.01~\AA$^{-1}$) for ARPES and 0.8 eV for XPS.
The crystals were cleaved {\it in situ} by the top-post method at 40~K under $\sim 7.0\times 10^{-11}$ Torr.
Surface adsorbates/intercalates were introduced on the sample surfaces by evaporating the pure (99.999\%) metal elements with a rate of 0.1~\AA/min in an ultra-high vacuum of $\sim$1.0$\times$10$^{-10}$~Torr at 50~K of the sample temperature.

\subsection{Acknowledgments}
This work was supported by the National Research Foundation (NRF) of Korea Grant funded by the Korean Government (MEST) (Nos. 2010-0010771 and 2013R1A1A2058195), by the NRF through the SRC Center for Topological Matter (No. 2011-0030046) and the Mid-Career Researcher Program (No. 2012-013838), and by the IBS in Korea.
The experiments at PLS were supported in part by MSIP and POSTECH.

\subsection{Competing financial interests}
The authors declare no competing financial interests.

\subsection{Author contributions}
The whole research was planned by H.J.N.
Bi$_2$Se$_3$ crystals were grown by J.P. and J.S.K.
ARPES/XPS measurements were carried out by J.J., I.K, E.J.C., B.G.P., H.D.K., and H.J.N.
H.J.N wrote the paper with suggestions and comments by H.D.K., E.J.C., and J.S.K.

\end{document}